# Interfacial Magnetic Vortex Formation in Exchange-Coupled Hard-Soft Magnetic Bilayers

X. H. Zhang[1,†,*], T. R. Gao,[1,†] L. Fang,[1] S. Fackler,[1] J. A. Borchers,[2] B. J. Kirby,[2] B. B. Maranville,[2] S. E. Lofland,[3] A. T. N'Diaye,[4] E. Arenholz,[4] A. Ullah,[5] J. Cui,[6,7] R. Skomski,[5] and I. Takeuchi[1,8*]

[1] Department of Materials Science and Engineering, University of Maryland, College Park, Maryland 20742, USA

[2] NIST Center for Neutron Research, National Institute of Standards and Technology, Gaithersburg, Maryland 20899, USA

[3] Department of Physics and Astronomy, Rowan University, Glassboro, NJ 08028, USA

[4] Advanced Light Source, Lawrence Berkeley National Laboratory, Berkeley, California 94720, USA

[5] Department of Physics and Astronomy and Nebraska Center for Materials and Nanoscience, University of Nebraska, Lincoln, Nebraska 68588, USA

[6] Division of Materials Science and Engineering, Ames Laboratory, Ames, Iowa 50011, USA

[7] Department of Materials Science and Engineering, Iowa State University, Ames, Iowa 50011, USA

[8] Quantum Materials Center, University of Maryland, College Park, Maryland 20742, USA

† These authors contributed equally to the work
* Email: xhzhang@umd.edu; takeuchi@umd.edu

**Abstract**

The exchange coupling between a hard magnetic layer MnBi and a soft magnetic layer Co-Fe has been found to significantly improve the maximum energy product. In this work, the spin structure of exchange-coupled MnBi:Co-Fe bilayers is experimentally investigated by X-ray magnetic circular dichroism (XMCD) and polarized neutron reflectometry (PNR). We find that the out-of-plane magnetization reversal process of the MnBi:Co-Fe bilayer structure involves formation of a curling-type twisting of the magnetization in the film plane at low or intermediate reversal fields. Micromagnetic simulations are further performed to provide a detailed view of the spins at the curling center. Reminiscent of chiral spin structures known as spin bobbers, this curling in the exchange-coupled hard-soft magnetic bilayers is a new type of skyrmionic spin structure and worth further investigation.

Manganese-bismuth (MnBi) is an intriguing alloy for several scientific and technological reasons. It is a magnetic model system characterized by complicated atomic [1,2,3], electronic [4,5], and magnetic [2,6,7] structures and with potential applications in permanent magnetism [8,9,10,11]. The technological fascination is fueled by the fact that both Mn and Bi are inexpensive elements, as compared to Pt and rare-earth metals. However, the magnetization of MnBi is relatively low, and attempts have been made to use hard-soft nanostructuring [12,13,14] to improve the energy product. Perpendicularly exchange-coupled MnBi:Co-Fe thin films, where MnBi is the hard phase and Co-Fe (cobalt-iron alloy) is the soft phase, are suitable for this purpose, and we have recently achieved an energy product enhancement from 98 kJ/m$^3$ to 172 kJ/m$^3$ in such a structure [11]. This is the highest nominal energy product achieved in a magnet without expensive or rare-earth elements.

The practical limitation of the hard-soft exchange-coupling approach is to maintain coercivity, that is, to avoid magnetization reversal in relatively small reverse fields [10,13]. This question has experimental and theoretical components: experimentally, the challenge is to avoid nucleation centers or to create pinning centers, which requires careful sample processing; theoretically, it is important to understand the magnetization reversal in the composite structure. The reversal starts in the soft phase and penetrates into the hard phase. Here, we show that our X-ray magnetic circular dichroism results, polarized neutron reflectometry results, and micromagnetic simulations all suggest that curling-type spin structures are formed during the magnetization reversal process.

At room temperature, the MnBi compound is in a low temperature phase (LTP) characterized by a hexagonal NiAs-like structure [1,15,16,17], and upon heating to 360 °C, bulk MnBi undergoes a structural transition into a high temperature phase (HTP) characterized by a Ni$_2$In-like structure [18]. The structural transition is accompanied by a magnetic transition from a ferromagnetic phase at low temperatures to an antiferromagnetic phase at high temperatures [19]. Strikingly, the temperature dependent coercivity of the LTP MnBi exhibits an anomalous increase with increasing temperature [20]. The energy product of MnBi and its composites is therefore expected to increase with temperature over a wide range [21].

To prepare the MnBi films on which the MnBi:Co-Fe bilayer structures are based on, a Bi layer and a Mn layer were sequentially deposited on a Si (100) substrate at room temperature in a magnetron sputtering system with a base pressure below $2.7 \times 10^{-6}$ Pa, and the bilayer structure was then post-annealed under high vacuum to form a Mn-Bi film. The Mn:Bi ratio of the film, tuned by controlling the thickness of each material layer during the deposition process, was quantitatively determined through wavelength dispersive X-ray spectroscopy measurements, and the microstructure of the films was studied by X-ray diffraction (XRD) using a diffractometer equipped with an area detector (Figure S1(a)). In addition, the hysteresis loop (Figure S1(b)) of each film was measured by a vibrating-sample magnetometer (VSM), and the film with a Mn:Bi ratio of 50:50 was found to show the highest perpendicular coercivity and the highest magnetization (Note: throughout this paper, the term MnBi refers to a Mn-Bi film with a Mn:Bi ratio of 50:50). Further, a thickness dependent study (Figure S1(c)) indicated that the equiatomic saturation magnetization of the MnBi films reaches a maximum value at a thickness of ~ 20 nm and an annealing-temperature study (Figure S1(d)) indicated that films annealed at temperatures equal to or slightly below 270 °C show the highest saturation magnetization. Please refer to Supplementary Materials for detailed characterization results of Mn-Bi films.

Since nanoscale exchange coupling is an attractive path to achieve large energy products [12,13,22,23], Co-Fe layers were deposited at room temperature on MnBi films, which were annealed at 270 ºC and cooled to room temperature under vacuum. Then, a Pt or $SiO_2$ layer with a thickness of 5 nm was deposited on the Co-Fe layer to prevent oxidation. As described in Supplementary Materials and in Ref. 11, we optimized the maximum energy product of the bilayer structure by changing the chemical composition of the Co-Fe alloy layer and/or changing the thicknesses of the soft layer and the hard layer respectively (Figure S2), and determined that the maximum energy product $(BH)_{max}$ of MnBi/Co-Fe bilayer structures can reach a value as high as 172 kJ/m$^3$.

Our previous polarized neutron reflectometry (PNR) experiments [11] have shown that after the magnetization of the bilayer structure is saturated in an out-of-plane direction, a moderate reverse field causes the magnetization to continuously rotate from the out-of-plane direction (i.e., the initial saturation direction) in the MnBi layer to the in-plane direction in the Co-Fe layer [11]. It is well-known that magnetization reversal of thin films with perpendicular magnetic anisotropy starts by magnetization curling in the film plane [24,25,26]. Curling is a vortex mode caused by magnetostatic self-interactions (flux closure) and represents an exact solution for some single-phase geometries. Our focus is on investigating the spin configurations in the MnBi/Co-Fe two-phase system by room-temperature X-ray magnetic circular dichroism (XMCD) and PNR [27,28,29], and by micromagnetic simulations using the *mumax* program package [30].

The XMCD and PNR measurements allow a layer-resolved analysis and the probing of magnetization components in different directions, and thus reflect the evolution of vortex states through a magnetization reversal process. Figure 1(a) shows both in-plane and out-of-plane hysteresis loops of a MnBi(20 nm)/Co$_{70}$Fe$_{30}$(3 nm) bilayer structure obtained from VSM measurements. A schematic cross-sectional view of the sample, including the Si substrate and the capping layer, is shown in the inset of Figure 1(a) and the X-ray reflectivity (XRR) result (Figure S3) confirms the thickness of each layer. Figures 1(b) and 1(c) show the respective element-specific XMCD hysteresis loops of Mn and Fe in the bilayer structure. The magnetic field is applied parallel to the polarized X-ray beam and perpendicular to the sample surface. The loops yield out-of-plane coercivities of 1.15 T for MnBi and 0.1 T for Fe, consistent with the VSM measurements. The coercivity of Fe (and Co, not shown here) is due to the exchange coupling between the MnBi and $Co_{70}Fe_{30}$ layers. Without this coupling, the coercivity of the Co-Fe layer would be virtually zero.

In Figure 1(d), in-plane XMCD hysteresis loops of Fe in the MnBi(20 nm)/$Co_{70}Fe_{30}$(5 nm) sample were obtained by applying a magnetic field in the film plane and using an X-ray beam grazed 30° to the film surface. The magnetization components $M_x$, $M_y$, and $M_z$ are parallel, transverse, and perpendicular to the magnetic-field direction, respectively. The "conventional" $M_x$ hysteresis loop indicates that the Fe is predominantly magnetized in the film plane, and the perpendicular magnetization component $M_z$ shows a typical hysteretic behavior in a field range between +100 mT and -100 mT. This confirms the exchange-coupling picture elaborated above. The small transverse in-plane component $M_y$ peaks near the coercivity field ($H = \pm H_C$) which, similar to that in other thin-film systems [31,32], is consistent with dynamic formation of chiral spin structures or possibly vortices by the Fe and Co moments during magnetization reversal in the $x$ direction.

In the PNR measurements, we independently measured the non-spin-flip (NSF) reflectivities $R^{++}$, $R^{--}$ and the spin-flip (SF) reflectivities $R^{+-}$, $R^{-+}$ as a function of magnetic field parallel to the sample surface after the sample was perpendicularly saturated in an off-line magnetic field of 1.6 T. In an NSF reflection process, the polarized neutrons do not change direction after interacting with the film. The NSF reflectivities are sensitive to the chemical compositional depth profile averaged across the sample plane, and the difference between $R^{++}$ and $R^{--}$ originates from the depth-dependent component of the in-plane magnetization parallel to the field. In the SF reflection process, the spin of a neutron rotates 180° after scattering from the sample, and the SF reflectivities are sensitive to the component of the in-plane magnetization perpendicular to the field. Together, the NSF and SF cross sections reveal the in-plane vector magnetization depth profile. In the in-plane field geometry, we measured the reflectivity of the MnBi(25 nm)/Co$_{70}$Fe$_{30}$(5 nm) bilayer structure at different magnetic fields incrementally from 0.004 T to 0.77 T. Figure 2 shows typical PNR results and corresponding fits (solid lines) for MnBi/Co$_{70}$Fe$_{30}$ with an in-plane field of 0.5 T.

Figure 3 shows the structural and magnetic depth profiles of the bilayer structure generated from the fits to the measured reflectivity data shown in Figure 2. The structural profile in Figure 3(a) matches that obtained from complementary PNR measurements in an out-of-plane field [11]. The fits indicate that the MnBi and Co$_{70}$Fe$_{30}$ thickness are 25 nm and 5 nm, respectively, though the boundary between them is smeared due to interlayer intermixing. In addition, the topmost section of the MnBi layer has reduced scattering length density (SLD) possibly indicative of voids or Mn-rich regions. Note that the structural depth profile is consistent with that obtained from X-ray reflectivity (XRR) measurements (Figure S4).

Figure 3(b) shows the corresponding depth-dependent profile of the in-plane magnetic SLD, which is directly proportional to the in-plane magnetization magnitude, at 0.5 T. The angle of the magnetization relative to the applied field (at 0°) as a function of depth is plotted in Figure 3(c). In a lower field of 0.004 T, the depth profile of the magnetization resembles that shown in Figure 3(b) for 0.5 T, but with a small in-plane magnetization mostly confined to the nominal Co-Fe layer. Upon increasing the in-plane field to 0.25 T and 0.5 T, Figure 3(b), a small component of the magnetization parallel to the in-plane field develops throughout the entire MnBi layer. This component emerges as a result of the application of the external field perpendicular to the out-of-plane *c*-axis of the MnBi (see, e.g., Refs. 8 and 33), the effect of which is somewhat enhanced by the exchange coupling to the Co-Fe layer.

The splitting in the NSF data along with the coexisting peak in the SF scattering near the critical angle is best fit (Figure 2) with a model that displays a twist of the in-plane magnetization throughout the Co-Fe layer. Specifically, the in-plane magnetization component at the bottom MnBi/Co-Fe interface is nearly parallel to the field but the magnetization component at the top Co-Fe/Pt interface is nearly perpendicular to the in-plane field, as shown in Figure 3(c) (Note that the magnitude of the in-plane transverse magnetization obtained from the fit may be smaller than its actual value since the spin flip scattering at low values of the wavevector $Q_z$ is shifted from the specular position due to the Zeeman effect [34]). The model used to fit the data in Figure 2 relies on the assumption that the sample vector magnetization is essentially homogenous across the sample plane. In this case, interpretation of the magnetic depth profile shown in Figures 3(b) and 3(c) is straightforward - the *net* magnetization averaged over the entire plane of the sample has an in-plane component that is twisted from the top of the MnBi layer through the CoFe layer. Such a twist could be the result of inherent differences in anisotropy between the two layers.

However, there is a plausible alternative interpretation of the data that does not require a net in-plane component of the magnetization perpendicular to the field across the entire sample. Specifically, incident neutron wavepackets scattered from individual magnetic regions with dimensions that exceed the neutron coherence length which is ≈100 μm parallel to the beam (*x* direction) and is ≈ 1 μm transverse to the beam (*y* direction) [35,36]. This results in an *incoherent* addition of scattering from different regions in the detector. Put differently, one would measure the average of different scattering patterns from distinct regions, rather than a single scattering pattern corresponding to the average in-plane structure. In this size regime, domains with oppositely-oriented perpendicular components would produce identical scattering patterns (Figure 4). Thus, an incoherent addition of scattering from those two types of domains would be virtually indistinguishable from that originating from an effectively single domain sample with a net in-plane magnetization component perpendicular to the field. In this alternate case, the twisting of the in-plane magnetization as a function of depth is consistent with the presence of magnetization curling. Since neutron reflectivity averages across the sample plane, these measurements would not be sensitive to the curling structure if the radius of the curling structure $R$ is less than the coherence length of the neutron. Figures 4(a) and 4(b), respectively, show a possible ordered spin structure that might form in small and moderate in-plane fields. The film plane (*x-y* plane) is the neutron scattering plane and the neutron wave vector $Q_z$ is parallel to the surface normal. The slits are narrow in the *x*-direction (along the beam direction) but wide in the *y*-direction (perpendicular to the beam path and parallel to the applied field), and the horizontal resolution (in the *x*-direction) is such that the neutron coherence length extends up to approximately 100 μm [35]. Specular $Q_z$ scans thus average coherently across the *x*-direction in the sample plane over distances that include multiple in-plane curling domains. The vertical resolution (*y*-direction) is coarse, and the neutron coherence length is only of the order of 0.1-0.5 μm. As a result, the $Q_z$ specular scan might actually be a composite of incoherent sums from regions along the *y*-axis of the sample that scatter independently, like individual samples.

In analogy to the reflectivity model for skyrmions in patterned multilayers [32], the sample in this model can be crudely divided into three regions (I - Curling domain top and surrounding sample, II - Curling domain bottom and surrounding sample, III - Region between domains, if any) as shown in Figure 4(a). The scattering from each region should add incoherently. For specular reflectivity scans, the difference in coherence lengths along the *x* and *y* directions means that the up/down spins on the sides of the curling domains in regions I and II will average to a net magnetization of zero if the applied field is small, as depicted in Figure 4 (a). However, the top and bottom magnetizations in the curling domains will give rise to SF scattering if the curling domain size is larger than the vertical ($y$) coherence length (< 1 µm). If the curling domains are random in-plane, rather than ordered, then the top and bottom magnetizations will generate SF scattering only if the domain size is greater than the horizontal ($x$) coherence length (approximately 100 µm).

The "up - down" symmetry, however, is broken within the sample plane in large in-plane fields which act to align the MnBi and CoFe magnetizations and to distort the curled structure. When a magnetic field $H$ is applied along the $y$-direction (represented by grey arrow), the "up" domains in the red area in Figure 4(b) will be expanded and the "down" domains will be shrunk in the green area. The distortion of the curling domain shown in Figure 4(b) will result in a net component of the magnetization (e.g., averaged over the coherence length along the $x$ direction) in the Co-Fe layer that is parallel to the in-plane field, and this component from regions I and II will contribute equally to the splitting of the NSF scattering that is evident in the 0.5 T reflectivity data in Figure 2. Since the MnBi/Co-Fe bilayer structure experimentally shows a very high out-of-plane coercivity, we can reasonably assume that the size of the curling domains within the layer is larger than the coherence length in the $y$-direction, in which case the top and bottom of each domain in regions I and II for this ordered structure should generate SF scattering of equal magnitude. Similar arguments again apply to a structure with curls that are disordered in-plane, though the curling domain size needs to exceed the large coherence length in the $x$-direction to produce SF scattering.

In practice, we fit the average "composite" structure and did not explicitly extract the characteristics of the magnetic structure in each of the three regions shown in Figure 4 to avoid over-parameterization since the sample exhibits significant interlayer intermixing. The depth profile of the vector magnetization [Figures 3(b) and 3(c)] suggests that the magnetization throughout the depth of the bilayer varies from an angle of almost (but not quite) 90° at the top of the Co-Fe layer to an angle of 0° relative to the field direction in the MnBi layer. By scattering symmetry, the data could be fit with magnetization that twists from an angle of almost (but not quite) 270° (-90°) at the top of the Co-Fe layer to an angle of 360° (0°) relative to the field in the MnBi layer. So, the fit is consistent with the expected scattering from **both** regions I and II, Figure 4(b), of the curling domain in which the Co-Fe magnetization winds through its depth toward the applied field direction, which corresponds to the direction of the in-plane component of the MnBi net magnetization. Since the expected reflectivity contributions from regions I and II are identical, however, we cannot definitively determine if the measured scattering originates from curling domains or just from randomly oriented in-plane domains with dimensions larger than the $x$ and $y$ neutron coherence lengths solely based on the PNR results. We note that our previous complementary measurements of the PNR investigations in an out-of-plane field (reported in Ref. 11) support either interpretation, though we would not expect to see any direct evidence of the curling twist in PNR measurements with the perpendicular measurement geometry.

As shown in Figure 4(a), the formation of in-plane curling domains in the Co-Fe layer *inevitably* leads to vertically-orientated spins at the center of each curling domain structure, suggesting that a chiral spin structure should be formed at the 'wind eye' of the curling domains. Within each chiral spin structure, the out-of-plane component of the spins in the Co-Fe layer gradually increases from the boundary to the center. Therefore, the magnetization of the Co-Fe layer has a finite out-of-plane component. As indicated in Figures 1(b) and 1(c), when an external magnetic field returns to zero after saturating the magnetization of the bilayer structure in an out-of-plane direction, the magnetization of the MnBi layer remains substantially along the out-of-plane direction, while the magnetization of the Co-Fe layer is reduced to a finite value in the out-of-plane direction, consistent with the formation of chiral spin structures. When a reversal field is applied, the out-of-plane component of the spins in the chiral spin structures starts to change the sign [11], and eventually leads to magnetization reversal of the MnBi layer and annihilation of the chiral spin structures at the out-of-plane coercivity field. This chiral spin structure formed in the exchange-coupled hard-soft magnetic bilayer is reminiscent of magnetic bobbers [37,38,39,40,41,42,43], which were discovered in the context of flux-closure domains [6,44] and have since attracted much attention, as exemplified by Refs. [8,45,33]. Magnetic bobbers are chiral in the sense of the spin texture topology [46], that is, the spins in a magnetic bobber can form left- or right-handed rotation structures. The current interest in magnetic bobbers is linked to Dzyaloshinskii-Moriya interactions in B20 [38] and related materials, which make these structures easier to stabilize on small length scales.

To further illustrate the formation of the chiral spin structures in the soft-hard magnetic bilayers, we perform micromagnetic simulations. The curling mode $\boldsymbol{M} = M_C(z, \rho)$ ($\cos\phi \boldsymbol{e}_y - \sin\phi \boldsymbol{e}_x$) originates from the competition between exchange energy and magnetostatic self-interaction [47,48,49,50,51]. The former favors coherent rotation, and the latter favors the partial flux closure inherent in the curling mode but costs exchange energy. Since exact solutions are available for simple systems only, it is convenient to use micromagnetic simulations. Figure 5 shows the curling-type spin structure for a MnBi(20 nm)/Co$_{70}$Fe$_{30}$(3 nm) thin-film patch in remanence after application of an out-of-plane saturation field. The simulations were performed using *mumax* [30]. The anisotropies of $K_1 = 0.9$ MJ/m$^3$ and $K_1 = 0$ for Co$_{70}$Fe$_{30}$ have been taken, and the computational cell size, equal to the distance between the arrows, is 1 nm in the *z*-direction and 2 nm in the *x*- and *y*-directions. The field is perpendicular to the film, and the spin-configuration snapshot $\mathbf{M}(\mathbf{r}, H_z)$ has been made close to coercivity, corresponding to the peaks in the red curve of Figure 1(d). It should be noted that in Figure 5, the simulation region of 40 nm × 40 nm (in the film plane) represents the "central point" of each curling domain structure illustrated in Figure 4.

Figure 5 confirms the main conclusions of the previous sections. In particular, it yields in-plane (or *x-y* plane) magnetization components that correspond to the magnetization component $M_y$ in Figure 1(d). Note that the curling mode is symmetric with respect to *x* and *y* in Figure 5, but not in Figure 1(d) because of the symmetry-breaking field applied in the *x*-direction. The chiral spin structure is primarily confined to the soft Co-Fe layer but somewhat penetrates into the hard MnBi layer. The curling intensity decays exponentially in the hard phase. The decay length of soft-into-hard penetration is generally of the order of $\delta_0 = \sqrt{A/K_1}$ [51], that is, about 3 nm for MnBi. This resistance to soft-hard penetration is the main reason for the high coercivities and energy products of the present MnBi/Co-Fe thin films. When the MnBi layer becomes too small, then two things happen. First, the film quality deteriorates due to the loss of Bi. Second, an unrelated magnetic effect is that the bottom part of the chiral spin structure starts to pierce the bottom of the MnBi layer, thereby facilitating magnetization reversal and causing coercivity and energy product to decrease.

Our calculations also show that the spin structure in Figure 5 carries a topological charge *Q*. This charge reflects the skyrmion-like noncollinearity of the spin structure and is obtained by integration over the skyrmion density $\Phi = \mathbf{M} \cdot (\partial \mathbf{M}/\partial x \times \partial \mathbf{M}/\partial y)/4\pi M_s^3$ [52]. The charge is field-dependent and reaches a sharp maximum in the chiral spin structure state near the coercivity, where the noncollinearities are most pronounced. In very high fields, $Q = 0$, because all spins are parallel and $\Phi = 0$. These spin structures are present at room temperature, due to the high $T_C$ values of MnBi and $Co_{70}Fe_{30}$. In addition, feature sizes in the spin structure are quite small, of the order of 10 nm. This is in the range of being potentially interesting in spin electronics [53]. One suggestion for future research is to investigate the topological charge experimentally. This can be done by Hall-effect measurements, because $Q$ corresponds to a Berry phase acquired by conduction electrons [54] and therefore to a topological Hall effect (THE) contribution [55] to the Hall resistivity. Another suggestion is to look at thin-film nanostructures made from other inversion-symmetric materials. Note that both MnBi and $Co_{70}Fe_{30}$ are macroscopically centrosymmetric, so no Dzyaloshinskii-Moriya interactions are necessary or involved in the formation of the chiral spin structures. In terms of Figure 5, right- and left-handed chiral spin structures are both possible and yield the same $Q$ and THE, irrespective of the chirality of the spin structure.

In conclusion, we have investigated the in-plane magnetization components in bilayers of *c*-axis-aligned hard MnBi and soft Co-Fe. These components play an important role in the understanding of the high coercivities and nominal energy products of MnBi:Co-Fe bilayers. Our VSM, XMCD and PNR experiments, combined with micromagnetic simulations, indicate that the magnetization has the character of topologically charged chiral spin structures. Compared to conventional spin bobbers, this new type of spin structure has a curling or vortex structure that touches the ground (the hard phase in the magnetic analogy) and partially penetrates it. Exchange-coupled nanostructures are therefore not only an option towards high-performing rare-earth-free permanent magnets but can also be used to create small-scale topological vortex structures at room temperature.

**Acknowledgement**

The work done at the University of Maryland was funded by the Department of Energy ARPA-E REACT (Grant No. 0472-1549), and the NIST Cooperative Agreement 70NANB17H301. The research at Nebraska is supported by the National Science Foundation under the award numbers EQUATE (OIA-2044049). We acknowledge the support of the National Institute of Standards and Technology, U.S. Department of Commerce, in providing access to the PBR and MAGIK reflectometers used in this work.

# References

[1] C. Guillaud, "Polymorphisme du composé défini Mn Bi aux températures de disparition et de réapparition de l'aimantation spontanée", J. Phys. Radium **12**, 143 (1951).

[2] J. B. Goodenough, *Magnetism and the Chemical Bond*, Wiley, New York 1963.

[3] A. Sarkar and A. B. Mallick, "Synthesizing the hard magnetic low-temperature phase of MnBi alloy: challenges and prospects", JOM **72**, 2812 (2020).

[4] R. Coehoorn and R. A. de Groot, "The electronic structure of MnBi", J. Phys. F **15**, 2135 (1985).

[5] A. Sakuma, Y. Manabe, and Y. Kota (2013), "First Principles Calculation of Magnetocrystalline Anisotropy Energy of MnBi and MnBi$_{1-x}$Sn$_x$", J. Phys. Soc. Jpn. **82**, 073704 (2013)

[6] B. W. Roberts and C. P. Bean, "Large Magnetic Kerr Rotation in BiMn Alloy", Phys. Rev. **96**, 1494-1496 (1954).

[7] P. Kharel, R. Skomski, P. Lukashev, R. Sabirianov, and D. J. Sellmyer, "Spin correlations and Kondo effect in a strong ferromagnet", Phys. Rev. B **84**, 014431 (2011).

[8] S. Chikazumi, *Physics of Magnetism*, Wiley, New York 1964.

[9] R. Skomski and J. M. D. Coey, *Permanent Magnetism*, Institute of Physics, Bristol 1999.

[10] R. Skomski, P. Manchanda, P. Kumar, B. Balamurugan, A. Kashyap, and D. J. Sellmyer, "Predicting the Future of Permanent-Magnet Materials" (invited), IEEE Trans. Magn. **49**, 3215 (2013).

[11] T. R. Gao, L. Fang, S. Fackler, S. Maruyama, X. H. Zhang, L. L. Wang, T. Rana, P. Manchanda, A. Kashyap, K. Janicka, A. L. Wysocki, A. T. N'Diaye, E. Arenholz, J. A. Borchers, B. J. Kirby, B. B. Maranville, K. W. Sun, M. J. Kramer, V. P. Antropov, D. D. Johnson, R. Skomski, J. Cui, and I. Takeuchi, "Large energy product enhancement in perpendicularly coupled MnBi/CoFe magnetic bilayers", Phys. Rev. B **94**, 060411(R) (2016).

[12] E. F. Kneller and R. Hawig, "The exchange-spring magnet: a new material principle for permanent magnets", IEEE Trans. Magn. **27**, 3588 (1991).

[13] R. Skomski and J. M. D. Coey, "Giant Energy Product in Nanostructured Two-Phase Magnets", Phys. Rev. B **48**, 15812 (1993).

[14] N. Jones, "The pull on stronger magnets", Nature **472**, 22 (2011).

[15] T. Chen and W. Stutius, "The phase transformation and physical properties of the MnBi and Mn$_{1.08}$Bi compounds", IEEE Trans. Magn. MAG. **10**, 581 (1974).

[16] J. B. Yang, W. B. Yelon, W. J. James, Q. Cai, S. Roy, and N. Ali, "Structure and magnetic properties of MnBi low temperature phase", J. Appl. Phys. **91**, 7866 (2002).

[17] T. SuwaT, Y. Tanaka, G. Mankey, R. Schad, and T. Suzuki, "Magnetic properties of low temperature phase MnBi of island structure", AIP Advances **6**, 056008 (2016).

[18] B. W. Roberts, "Neutron diffraction study of the structures and magnetic properties of Manganese Bismuthide", Phys. Rev. **104**, 607 (1956).

[19] R. R. Heikes, "Magnetic transformation in MnBi", Phys. Rev. **99**, 446 (1955).

[20] J. B. Yang, Y. B. Yang, X. G. Chen, X. B. Ma, J. Z. Han, Y. C. Yang, S. Guo, A. R. Yan, Q. Z. Huang, M. M. Wu, and D. F. Chen, "Anisotropic nanocrystalline MnBi with high coercivity at high temperature", Appl. Phys. Lett. **99**, 082505 (2011).

[21] S. Saha, R. T. Obermyer, B. J. Zande, V. K. Chandhok, S. Simizu, and S. G. Sankar, "Magnetic properties of the low-temperature phase of MnBi", J. Appl. Phys. **91**, 8525 (2002).

[22] Y. Q. Li, M. Yue, T. Wang, Q. Wu, D. T. Zhang, and Y. Gao, "Investigation of magnetic properties of MnBi/Co and MnBi/Fe$_{65}$Co$_{35}$ Nanocomposite permanent magnets by micro-magnetic simulation", J. Magn. Magn. Mater. **393**, 484 (2015).

[23] Y. L. Ma, X. B. Liu, K. Gandha, N. V. Vuong, Y. B. Yang, J. B. Yang, N. Poudyal, J. Cui, and J. P. Liu, "Preparation and magnetic properties of MnBi-based hard/soft composite magnets", J. Appl. Phys. **115**, 17A755 (2014).

[24] W. F. Brown, "Criterion for uniform micromagnetization", Phys. Rev. **105**, 1479 (1957) and "Virtues and weaknesses of the domain concept", Rev. Mod. Phys. **17**, 15 (1945).

[25] A. Aharoni, *Introduction to the Theory of Ferromagnetism*, University Press, Oxford 1996.

[26] R. Skomski, H.-P. Oepen, and J. Kirschner, "Micromagnetics of ultrathin films with perpendicular magnetic anisotropy", Phys. Rev. B **58**, 3223 (1998).

[27] C. F. Majkrzak, K. V. O'Donovan, and N. F. Berk, *Neutron Scattering from Magnetic Materials*, Elsevier Science, New York 2005.

[28] M. R. Fitzsimmons and C. F. Majkrzak, *Application of Polarized Neutron Reflectometry to Studies of Artificially Structured Magnetic Materials, in Modern Techniques for Characterizing Magnetic Materials*, Springer, Berlin 2005.

[29] B. J. Kirby, P. A. Kienzle, B. B. Maranville, N. F. Berk, J. Krycka, F. Heinrich, C. F. Majkrzak, "Phase-sensitive specular neutron reflectometry for imaging the nanometer scale composition depth profile of thin-film materials", Current Opinion in Colloid and Interface Science **17**, 44 (2012).

[30] A. Vansteenkiste, J. Leliaert, M. Dvornik, M. Helsen, F. Garcia-Sanchez, and B. van Waeyenberge, "The design and verification of mumax3", AIP Advances **4**, 107133 (2014).

[31] K. S. Buchanan, K. Yu. Guslienko, A. Doran, A. Scholl, S. D. Bader, and V. Novosad, "Magnetic remanent states and magnetization reversal in patterned trilayer nanodots", Phys. Rev. B **72**, 134415 (2005).

[32] D. A. Gilbert, B. B. Maranville, A. L. Balk, B. J. Kirby, P. Fischer, D. T. Pierce, J. Unguris, J. A. Borchers, and K. Liu, "Realization of ground-state artificial Skyrmion lattices at room temperature", Nat. Commun. **6**, 8462 (2015).

[33] R. Skomski, *Simple Models of Magnetism*, University Press, Oxford 2008.

[34] B. B. Maranville, B. J. Kirby, A. J. Grutter, P. A. Kienzle, C. F. Majkrzak, Y. Liu, and C. L. Dennis, "Measurement and modeling of polarized specular neutron reflectivity in large magnetic fields", J. Appl. Cryst. **49**, 1121 (2016).

[35] C. F. Majkrzak, C. Metting, B. B. Maranville, J. A. Dura, S. Satija, T. Udovic, and N. F. Berk, "Determination of the effective transverse coherence of the neutron wave packet as employed in reflectivity investigations of condensed-matter structures. I. Measurements", Phys. Rev. A **89**, 033851 (2014).

[36] L. Fallarino, B. J. Kirby, M. Pancaldi, P. Riego, A. L. Balk, C. W. Miller, P. Vavassori, and A. Berger, Phys. Rev. B **95**, 134445 (2017).

[37] F. N. Rybakov, A.B. Borisov, S. Blügel, and N. S. Kiselev, "New Type of Stable Particlelike States in Chiral Magnets", Phys. Rev. Lett. **115**, 117201 (2015).

[38] F. Zheng, F. N. Rybakov, A. B. Borisov, D. Song, Sh. Wang, Z.-A. Li, H. Du, N. S. Kiselev, J. Caron, A. Kovács, M. Tian, Y. Zhang, S. Blügel, and R. E. Dunin-Borkowski, "Experimental observation of chiral magnetic bobbers in B20-type FeGe", Nature Nanotechnology **13**, 451 (2018).

[39] S. L. Zhang, G. van der Laan, W. W. Wang, A. A. Haghighirad, and T. Hesjedal, "Direct Observation of Twisted Surface skyrmions in Bulk Crystals", Phys. Rev. Lett. **120**, 227202 (2018).

[40] A. O. Leonov and K. Inoue, "Homogeneous and heterogeneous nucleation of skyrmions in thin layers of cubic helimagnets", Phys. Rev. B **98**, 054404 (2018).

[41] M. Redies, F. R. Lux, J.-P. Hanke, P. M. Buhl, G. P. Müller, N. S. Kiselev, S. Blügel, and Y. Mokrousov, "Distinct magnetotransport and orbital fingerprints of chiral bobbers", Phys. Rev. B **99**, 140407(R) (2019).

[42] A. S. Ahmed, J. Rowland, B. D. Esser, S. R. Dunsiger, D. W. McComb, M. Randeria, and R. K. Kawakami, "Chiral bobbers and skyrmions in epitaxial FeGe/Si(111) films", Phys. Rev. Materials **2**, 041401(R) (2018).

[43] K. Ran, Y. Liu, Y. Guang, D. M. Burn, G. van der Laan, T. Hesjedal, H. Du, G. Yu, and Sh. Zhang, "Creation of a Chiral Bobber Lattice in Helimagnet-Multilayer Heterostructures", Phys. Rev. Lett. **126**, 017204 (2021).

[44] Y. Takata, "Observation of Domain Structure and Calculation of Magnetostatic Energy on the c-Plane of Cobalt Single Crystals", J. Phys. Soc. Jpn. **18**, 87 (1963).

[45] A. Hubert and R. Schäfer, *Magnetic Domains*, Springer-Verlag, Berlin 1998.

[46] P. J. Grundy and S. R. Herd, "Lorentz Microscopy of Bubble Domains and Changes in Domain Wall State in Hexaferrites", phys. stat. sol. (a) **20**, 295 (1973).

[47] S. K. Sharma, H. R. Prakash, S. Ram, and D. Pradhan, "Synthesis and Magnetic Properties of Rare-Earth Free MnBi Alloy: A High-Energy Hard Magnetic Material", AIP Conference Proceedings **1942**, 130044 (2018).

[48] N. Poudyal, X. Liu, W. Wang, V. V. Nguyen, Y. Ma, K. Gandha, K. Elkins, J. P. Liu, K. Sun, M. J. Kramer, and J. Cui, "Processing of MnBi Bulk Magnets with Enhanced Energy Product", AIP Advances **6**, 056004 (2016).

[49] R. Skomski, P. Manchanda, I. Takeuchi, and J. Cui, "Geometry dependence of magnetization reversal in nanocomposite alloys", JOM **66**, 1144 (2014).

[50] T. H. Rana, P. Manchanda, B. Balamurugan, A. Kashyap, T. R. Gao, I. Takeuchi, J. Cui, S. Biswas, R. F. Sabirianov, D. J. Sellmyer, "Micromagnetism of MnBi:FeCo thin films", J. Phys. D: Appl. Phys. **49**, 075003 (2016).

[51] R. Skomski, "Nanomagentics", J. Phys.: Condens. Matter **15**, R841 (2003).

[52] Sh. Seki and M. Mochizuki, *Skyrmions in Magnetic Materials*, Springer International, Cham 2016.

[53] D. Xiao, M.-Ch. Chang, and Q. Niu, "Berry phase effects on electronic properties", Rev. Mod. Phys. **82**, 1959 (2010).

[54] B. Balasubramanian, P. Manchanda, R. Pahari, Zh. Chen, W. Zhang, S. R. Valloppilly, X. Li, A. Sarella, L. Yue, A. Ullah, P. Dev, D. A. Muller, R. Skomski, G. C. Hadjipanayis, and D. J. Sellmyer, "Chiral magnetism and high-temperature skyrmions in B20-ordered Co-Si", Phys. Rev. Lett. **124**, 057201 (2020).

[55] Y. Aharonov and A. Stern, "Origin of the geometric forces accompanying Berry's geometric potentials, Phys. Rev. Lett. **69**, 3593 (1992).

**Figure Captions**

Figure 1. Hysteresis loops of a MnBi(20 nm)/Co$_{70}$Fe$_{30}$(3 nm) bilayer sample. (a) the in-plane and out-of-plane hysteresis loops obtained by VSM; and element-sensitive room-temperature XMCD hysteresis loops: (b) out-of-plane loop for Mn, (c) out-of-plane loop for Fe, and (d) in-plane loops for Fe with $M_x$ (black), $M_y$ (red) and $M_z$ (blue) components. Note that the scale of $M_y$ and $M_z$ is magnified by a factor of 10.

Figure 2. Polarized neutron reflectivity data and corresponding fits (solid lines) for all four cross sections, R++ (blue), R+- (purple), R-+ (green) and R-- (red) obtained in a field of 0.5 T applied parallel to the sample plane of a MnBi(25 nm)/Co$_{70}$Fe$_{30}$(5 nm) bilayer after saturation in a perpendicular field of 1.6 T. Note: error bars displayed here indicate 1 standard deviation.

Figure 3. Depth profiles of a MnBi(25 nm)/Co$_{70}$Fe$_{30}$(5 nm) bilayer: (a) structural scattering length density (SLD), (b) magnetic scattering length density which is proportional to the in-plane component of the magnetization, and (c) angle of the in-plane magnetization component relative to the 0.5 T in-plane field at 0° obtained from fits to polarized neutron reflectometry measurements. The rotation of the in-plane magnetization as a function of depth from 0° near the MnBi/Co-Fe interface to 90° near the Co-Fe/Pt interface is consistent with presence of the curling mode.

Figure 4. Cartoon of magnetization reversal with curling model in the topmost of Co$_{70}$Fe$_{30}$ layer (X-Y plane) at initial state in a low field (a) and after applying a large field (b). The white arrows represent magnetization, and the grey arrows represent the strength and the direction of the magnetic field. Each of the three regions in (a) and (b) scatter coherently, and the resultant reflectivity is an incoherent sum of the scattering from each region.

Figure 5. Magnetization curling in MnBi(20 nm)/Co$_{70}$Fe$_{30}$(3 nm) determined from micromagnetic calculations. A 40 nm × 40 nm area is shown in the figure as an enlarged view of a "wind eye" where red, blue, green, and yellow regions meet in Figure 7. The arrows represent orientations of local spins, and the colors (i.e., red, blue, green, and yellow) represent the local in-plane magnetization direction (consistent with Figure 7). The top-most region is in plane as it is the Co$_{70}$Fe$_{30}$ layer, and then it gradually changes to out of plane in the depth direction as it transitions to the MnBi layer.

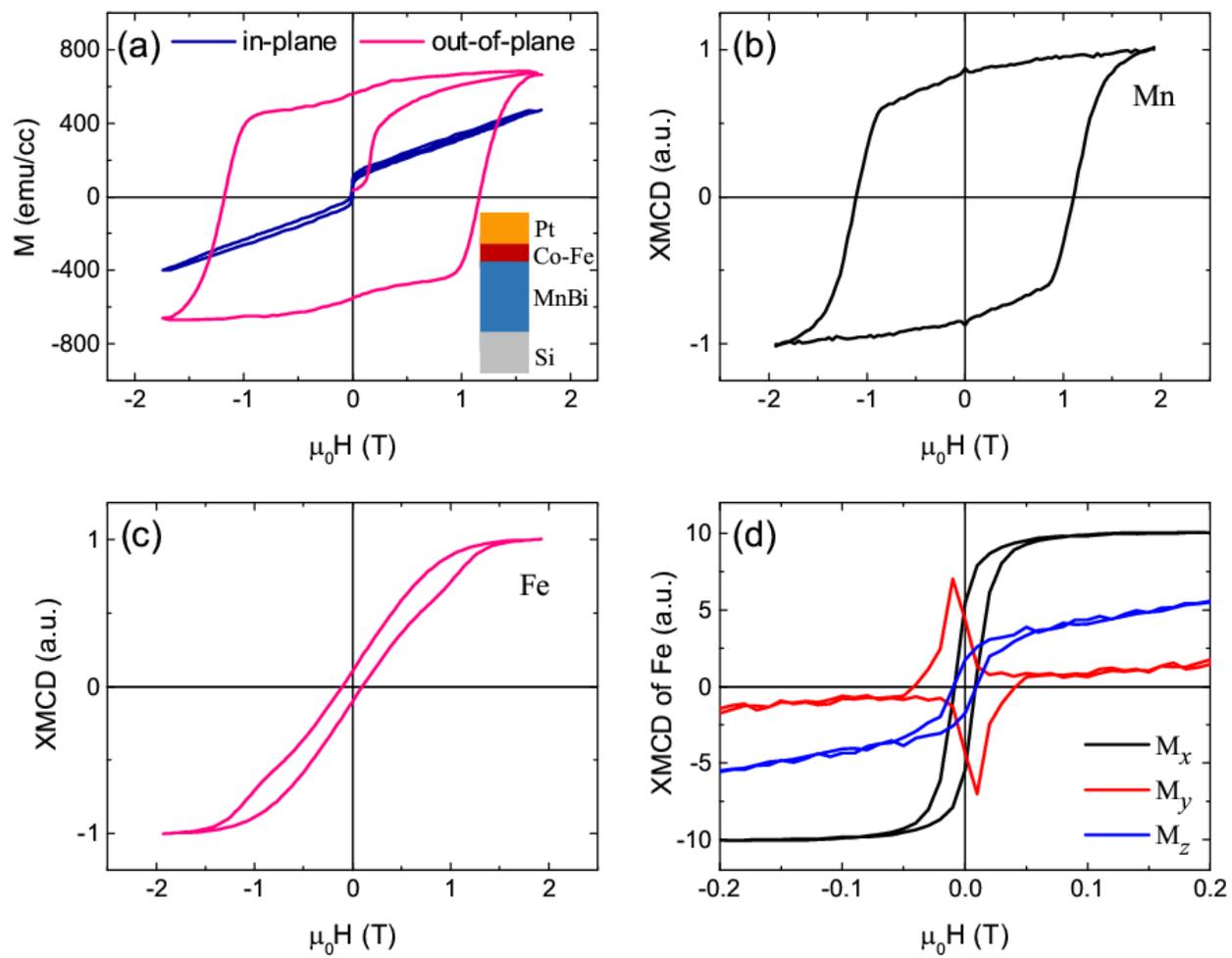

Figure 1

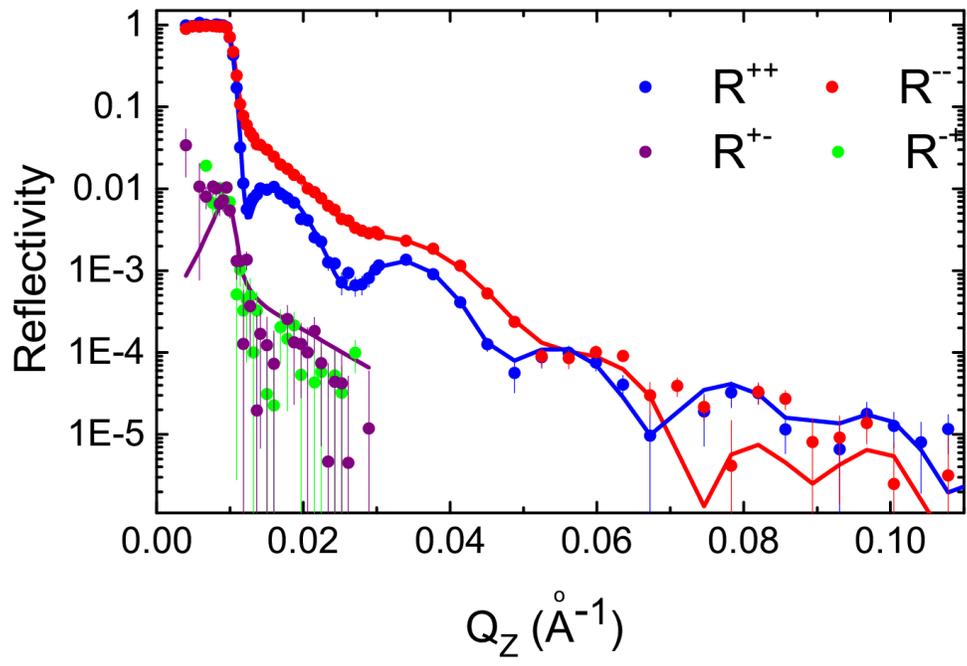

Figure 2

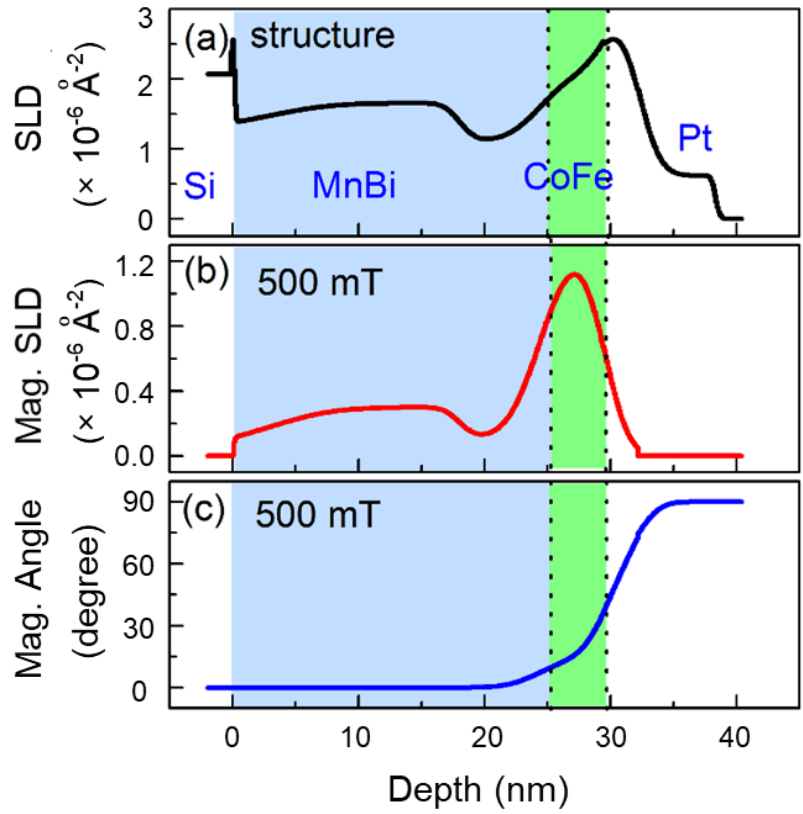

Figure 3

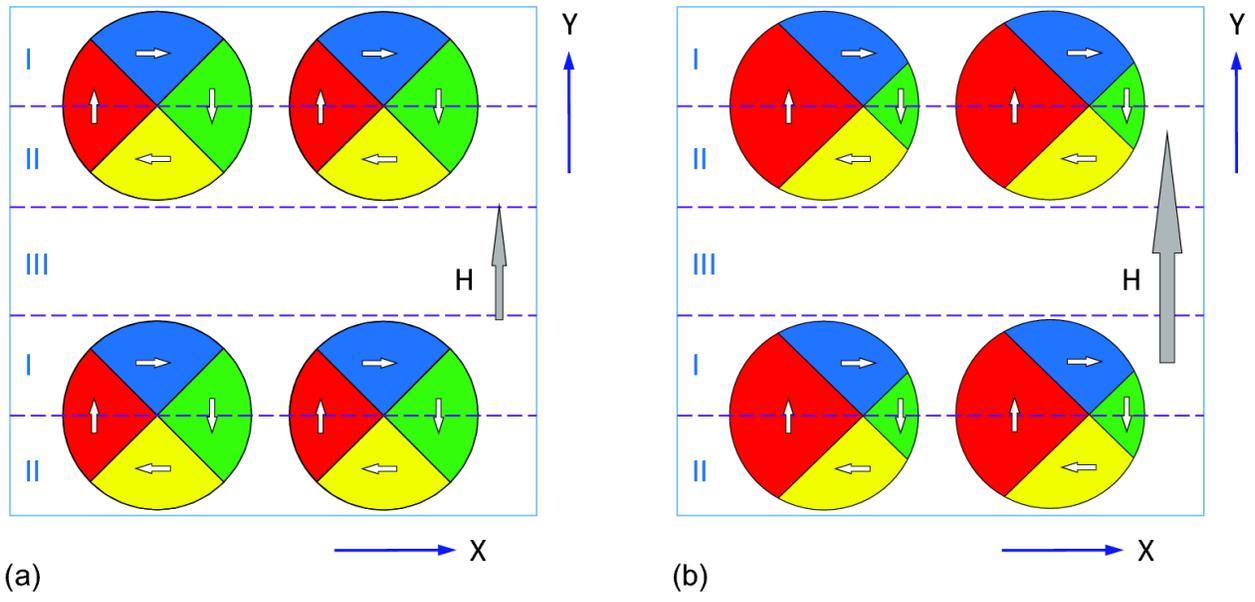

Figure 4

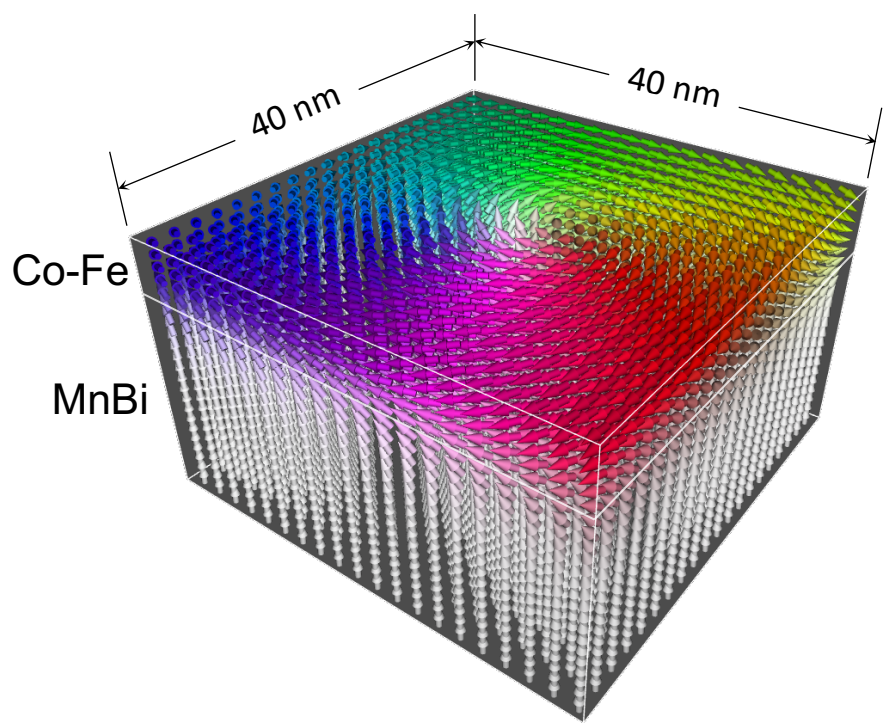

Figure 5

# Interfacial Magnetic Vortex Formation in Exchange-Coupled Hard-Soft Magnetic Bilayers

# Supplementary Materials

X. H. Zhang[1,†,*], T. R. Gao,[1,†] L. Fang,[1] S. Fackler,[1] J. A. Borchers,[2] B. J. Kirby,[2] B. B. Maranville,[2] S. E. Lofland,[3] A. T. N'Diaye,[4] E. Arenholz,[4] A. Ullah,[5] J. Cui,[6,7] R. Skomski,[5] and I. Takeuchi[1,8*]

[1] Department of Materials Science and Engineering, University of Maryland, College Park, Maryland 20742, USA

[2] NIST Center for Neutron Research, National Institute of Standards and Technology, Gaithersburg, Maryland 20899, USA

[3] Department of Physics and Astronomy, Rowan University, Glassboro, NJ 08028, USA

[4] Advanced Light Source, Lawrence Berkeley National Laboratory, Berkeley, California 94720, USA

[5] Department of Physics and Astronomy and Nebraska Center for Materials and Nanoscience, University of Nebraska, Lincoln, Nebraska 68588, USA

[6] Division of Materials Science and Engineering, Ames Laboratory, Ames, Iowa 50011, USA

[7] Department of Materials Science and Engineering, Iowa State University, Ames, Iowa 50011, USA

[8] Quantum Materials Center, University of Maryland, College Park, Maryland 20742, USA

† These authors contributed equally to the work

* Email: xhzhang@umd.edu; takeuchi@umd.edu

1. **Characterization of the properties of Mn-Bi thin films**

Figure S1(a) shows an XRD pattern of a MnBi (100 nm) film measured at room temperature. The (002) ($2\theta = 29.1°$) and (004) ($2\theta = 60.3°$) diffraction peaks of MnBi show relatively high intensities, suggesting that the film has a strong hexagonal LTP NiAs structure with preferential orientation along the *c*-axis. It should be noted that in addition to the (002) and (004) diffraction peaks, some other peaks of MnBi, e.g., the (102) peak with $2\theta \sim 38.0°$ and the (202) peak with $2\theta \sim 58.0°$, can also be identified from the pattern. Moreover, the pattern also shows several Bi peaks, which is consistent with the reported inevitable presence of Bi in MnBi magnetic alloys [S47,S48,S3].

The hysteresis loops of the Mn-Bi films, measured by a vibrating-sample magnetometer (VSM), strongly depend on synthesis conditions such as the film thickness, the Mn to Bi ratio, annealing temperature, and annealing time. Throughout this paper, the film plane is in an *x-y* plane, the thickness of the film is in a *z*-direction, and the external magnetic field is applied in the *x*-direction and/or in the *z*-direction, depending on the experiment. Figure S1(b) shows typical out-of-plane hysteresis loops for different Mn:Bi ratios. The films shown in this figure have a thickness of 20 nm and are annealed at 270 °C. With a nominal Mn:Bi ratio of 50:50, the MnBi film shows the highest perpendicular coercivity (1.6 T) and the highest magnetization (660 kA/m). However, consistent with the Brown's coercivity paradox [S24,S25,S6], this highest coercivity is still much smaller than the anisotropy field of MnBi. Please note that throughout this paper, the term MnBi refers to films with a Mn:Bi ratio of 50:50, and films with other Mn:Bi ratios are only prepared for the VSM study shown in Figure S1(b).

Figure S1(c) shows the thickness dependence of the saturation magnetization of the MnBi films post-annealed at 270 °C. The magnetization increases as the film thickness increases from 10 nm to 20 nm, where it reaches a maximum. The reason for the comparatively low magnetization of very thin films is very likely the degradation MnBi due to the volatile nature of Bi. Above 20 nm, the Mn and Bi layers may not be able to react completely and form a chemically uniform MnBi film, which explains the magnetization reduction in thicker films. The saturation magnetization of the 20 nm MnBi film increases as the annealing temperature increases from 180 °C to about 210 °C, as shown in Figure S1(d). The saturation magnetization remains nearly constant in the temperature range from 230 °C to 270 °C, and then shows a relatively sharp decrease when the annealing temperature exceeds 270 °C.

## 2. Structural optimization for MnBi/Co-Fe bilayers

Figure S2(a) shows an out-of-plane hysteresis loop of a MnBi(20 nm)/Co(5 nm) bilayer structure at room temperature, and from this exemplary hysteresis loop, a maximum energy product $(BH)_{max}$ of the bilayer structure can be obtained. As shown in Figure S2(b), in MnBi/Co bilayer structures, our previous study of the nominal maximum energy product $(BH)_{max}$ as a function of the Co thickness (with a fixed MnBi thickness of 20 nm) [S11] indicates that the maximum energy product $(BH)_{max}$ reaches the highest value when the thickness of the Co layer is 3 nm. The strong dependence of $(BH)_{max}$ on the Co layer thickness reflects the subtle interplay between the exchange energy and magnetostatic self-interaction in the chiral domains. Figure S2(c) shows $(BH)_{max}$ as a function of MnBi thickness in MnBi/Co bilayer structures with a fixed Co thickness of 3 nm. The nominal maximum energy product $(BH)_{max}$ shows an initial increase as the MnBi thickness increases and reaches 172 kJ/m$^3$ when the thickness of the MnBi layer is 20 nm. This thickness corresponds to the penetration depth of the bobber tip into the hard phase determined from numerical simulations. A thinner MnBi layer is fully penetrated by the bobber tip, and thus negatively affects the coercivity and energy product. A detailed discussion of this point is included in the numerical simulation section. As shown in Figure S2(c), further increasing the MnBi thickness to exceed 20 nm reduces the maximum energy product because the magnetization per volume is reduced for the MnBi layer.

3. **XRR results obtained from the sample used in the XMCD measurements**

Figure S3 shows the X-ray reflectivity (XRR) results (red open circles) obtained from the MnBi(20 nm)/Co$_{70}$Fe$_{30}$(3 nm) bilayer sample that was studied in the XMCD measurements. The best fit (blue line) to the data results in a MnBi thickness of 25.1 nm, a Co-Fe thickness of 3.2 nm, and a Pt thickness of 4.0 nm.

4. **XRR results obtained from the sample used in the PNR measurements**

Figure S4 shows the XRR results (blue dots) obtained from the MnBi(25 nm)/Co$_{70}$Fe$_{30}$(5 nm) bilayer sample that was studied in the PNR measurements. The best fit (blue line) to the data results in a MnBi thickness of 26.3 nm, a Co-Fe thickness of 4.4 nm, and a Pt thickness of 5.8 nm.

**References**

[S56] S. K. Sharma, H. R. Prakash, S. Ram, and D. Pradhan, "Synthesis and Magnetic Properties of Rare-Earth Free MnBi Alloy: A High-Energy Hard Magnetic Material", AIP Conference Proceedings **1942**, 130044 (2018).
[S57] N. Poudyal, X. Liu, W. Wang, V. V. Nguyen, Y. Ma, K. Gandha, K. Elkins, J. P. Liu, K. Sun, M. J. Kramer, and J. Cui, "Processing of MnBi Bulk Magnets with Enhanced Energy Product", AIP Advances **6**, 056004 (2016).
[S58] C. Li, D. Guo, B. Shao, K. Li, B. Li, and D. Chen, "Effect of Heat Treatment and Ball Milling on MnBi Magnetic Materials", Mater. Res. Express **5**, 016104 (2018).

[S59] W. F. Brown, "Criterion for uniform micromagnetization", Phys. Rev. **105**, 1479 (1957) and "Virtues and weaknesses of the domain concept", Rev. Mod. Phys. **17**, 15 (1945).

[S60] A. Aharoni, *Introduction to the Theory of Ferromagnetism*, University Press, Oxford 1996.

[S61] B. Balasubramanian, P. Mukherjee, R. Skomski, P. Manchanda, B. Das, and D. J. Sellmyer, "Magnetic nanostructuring and overcoming Brown's paradox to realize extraordinary high-temperature energy products", Sci. Rep. **4**, 6265 (2014).

[S62] T. R. Gao, L. Fang, S. Fackler, S. Maruyama, X. H. Zhang, L. L. Wang, T. Rana, P. Manchanda, A. Kashyap, K. Janicka, A. L. Wysocki, A. T. N'Diaye, E. Arenholz, J. A. Borchers, B. J. Kirby, B. B. Maranville, K. W. Sun, M. J. Kramer, V. P. Antropov, D. D. Johnson, R. Skomski, J. Cui, and I. Takeuchi, "Large energy product enhancement in perpendicularly coupled MnBi/CoFe magnetic bilayers", Phys. Rev. B **94**, 060411(R) (2016).

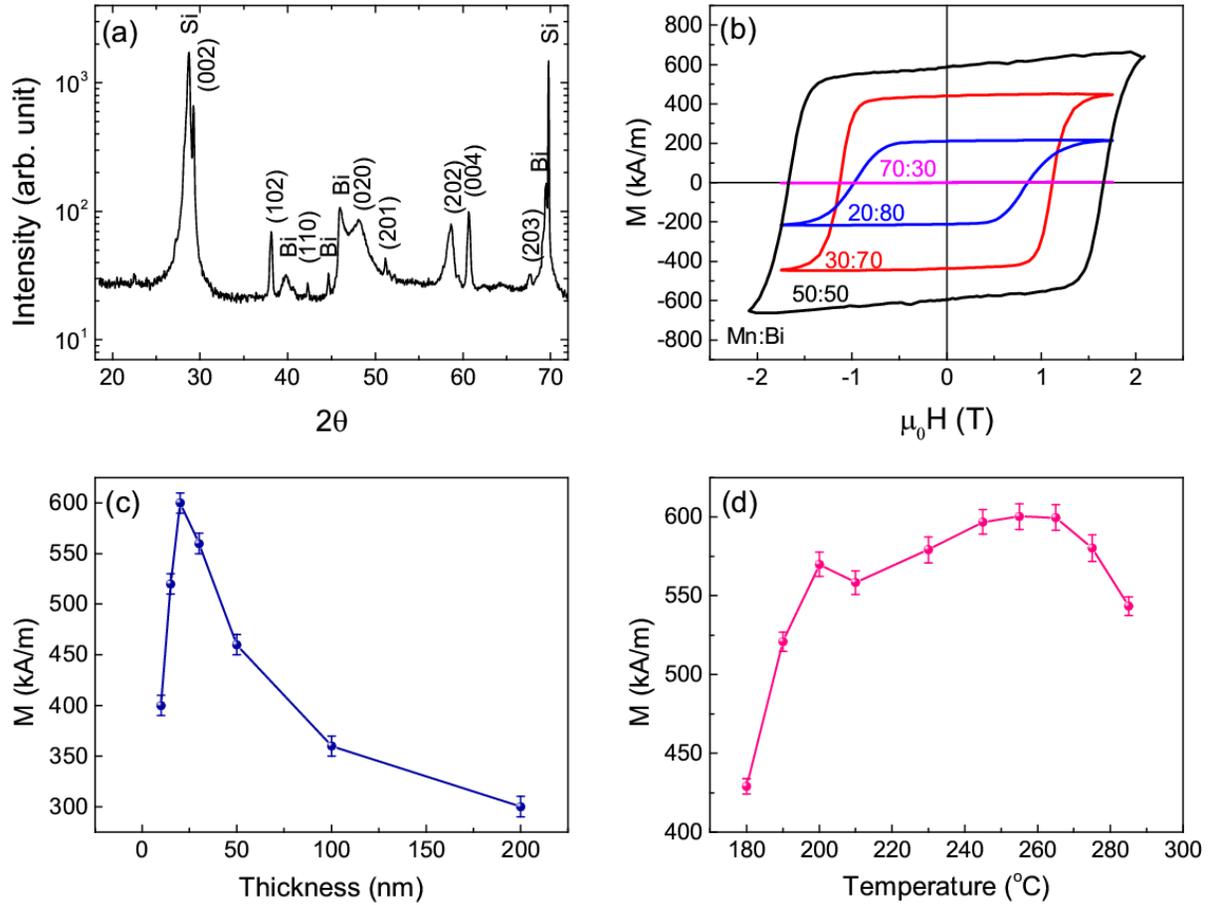

Figure S1. Structural and magnetic properties of the Mn-Bi films: (a) XRD pattern of a MnBi (100 nm) film at room temperature; (b) hysteresis loops of 20 nm films with different atomic concentrations annealed at 270 ºC, (c) thickness dependence of the saturation magnetization of equiatomic MnBi films annealed at 270 ºC, and (d) magnetization of equiatomic MnBi films as a function of annealing temperature. In (b), the atomic ratios of Mn to Bi are 50:50, 30:70, 20:80, and 70:30, respectively for the films, and the magnetization is measured in the *z*-direction perpendicular to the film. Note: error bars displayed in (c) and (d) indicate 2 standard deviations.

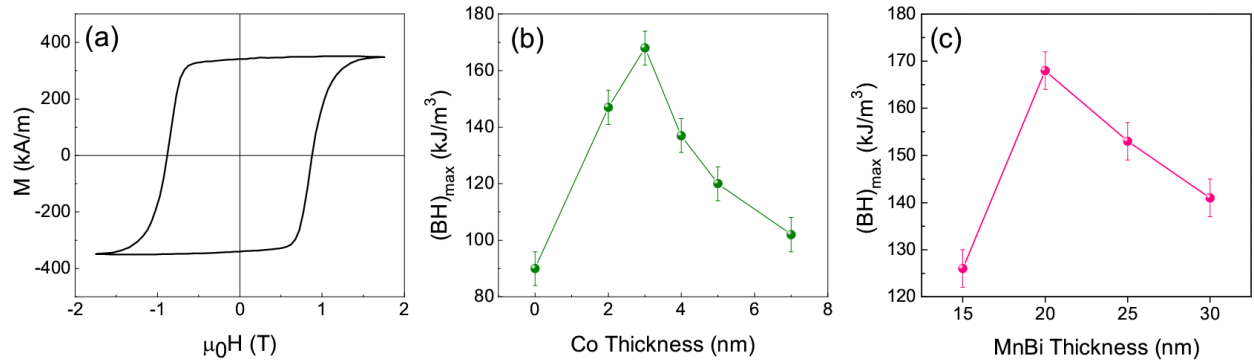

Figure S2. Hysteresis of MnBi/Co bilayers: (a) an out-of-plane hysteresis loop of a MnBi(20 nm)/Co(5 nm) bilayer structure at room temperature, (b) and (c) energy product as a function of Co thickness (with a fixed MnBi thickness of 20 nm) and MnBi thickness (with a fixed Co thickness of 3 nm), respectively. Note: (b) is reproduced from ref. 62, and error bars displayed in (b) and (c) indicate 2 standard deviations.

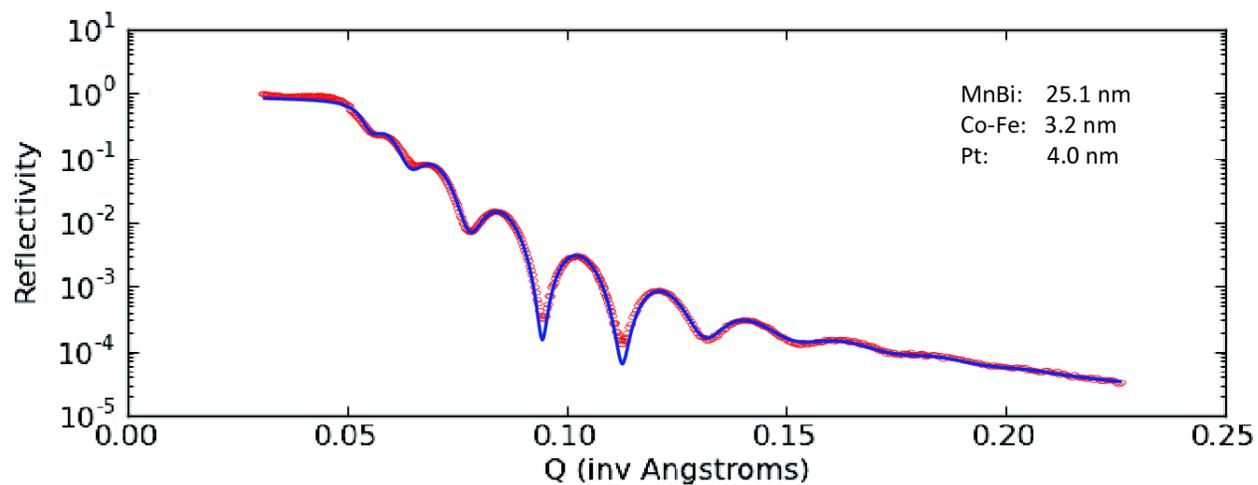

Figure S3. XRR results (red open circles) of a MnBi(20 nm)/Co$_{70}$Fe$_{30}$(3 nm) bilayer sample and the best fit (blue line) to the data. The sample was used in the XMCD measurements.

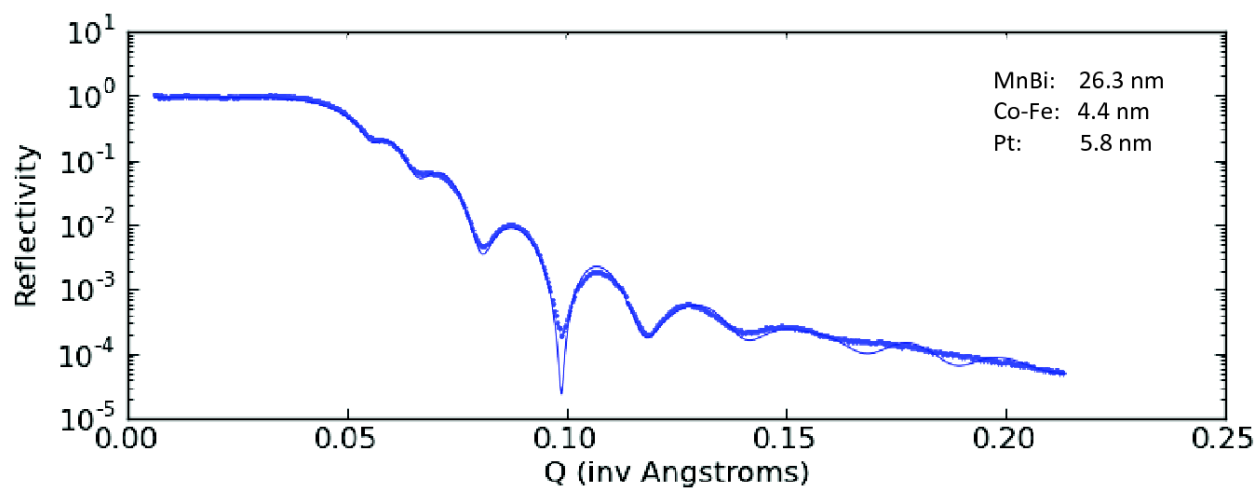

Figure S4. XRR results (blue dots) of a MnBi(25 nm)/Co$_{70}$Fe$_{30}$(5 nm) bilayer sample and the best fit (blue line) to the data. The sample was used in the PNR measurements.